\documentclass[preprint,letter]{ptptex}
\usepackage[T1]{fontenc}
\usepackage[latin9]{inputenc}
\usepackage{amstext}
\usepackage{amssymb}

\makeatletter

\usepackage{pxfonts}

\title{
Relativistic Statistical Mechanics with Angular Momentum
}

\author{
Tadas K. \textsc{Nakamura}%
}

\inst{
Fukui Prefectural Universiyt, Fukui 910-1195, Japan
}

\abst{
The equilibrium distribution function of a relativistic ideal gas has
been derived to include the effect of angular momentum.  The result
agrees with the one obtained from kinetic theory, and consistent with
relativistic thermodynamics. The role of angular momentum becomes
transparent in this derivation, and the equilibrium distribution can
be generalized to accommodate the effect of intrinsic angular
momentum. The results here are for flat spacetimes, however, same
approach can be applied to static curved spacetimes.}

\begin{document}

\maketitle

The first law of thermodynamics is energy conservation. Energy plays
the fundamental role in thermodynamics because of its conservation,
and temperature is defined corresponding to thermal energy. Provided
that the conservation law causes the temperature for energy, then
other conserved quantities can result in thermodynamical parameters
similar to the temperature. For example, chemical potential is
introduced corresponding to the conservation of particle
number. 

However, momentum, yet another conserved quantity, is rarely
considered in thermodynamics because to include it is trivial
generalization with Gallilei transform in non-relativistic dynamics;
energy can be decomposed into a simple sum of thermal energy and energy
of bulk motion. However, such decomposition is not covariant in
relativity, and thus momentum has non-trivial effects in relativistic
thermodynamics.

It has been proposed to define an inverse temperature corresponding to
each component of momentum, and its three components form a four
vector together with the inverse temperature of energy
\cite{kampen68PhRv,israel76AnPhy}.  This four vector is called inverse
temperature four vector, and with it we can construct manifestly
covariant relativistic thermodynamics. The inverse temperature four
vector is first introduced for a body with finite volumes in
equilibrium \cite{kampen68PhRv,israel76AnPhy,tadas06PhLA}, and it is
widely accepted in the relativistic thermodynamics of a continuum
(see, e.g., Maartens\cite{maartens1996arXiv}).

Provided that momentum has a role similar to energy because of its
conservation, angular momentum can be treated in the same way as a
conserved quantity. Recently Nakamura
\cite{nakamura_thermodynamics_2010} has formulated relativistic
thermodynamics with angular momentum based on this idea. Inverse
temperature that corresponds to angular momentum is introduced as a
covariant bi-vector (anti-symmetric rank two tensor) because angular
momentum is a contravariant bi-vector in relativity. The purpose of
the present paper is to derive this inverse temperature bi-vector from
the view point of statistical mechanics.

\bigskip{}

We seek for the equilibrium distribution by maximizing the entropy
under the constraints of conservation laws. The foundation of this
approach may be highly conceptual and unsolved enigma in statistical
mechanics. We may attribute its basis to the ergodic assumption of
conventional statistical mechanics, or maximum entropy principle from
the information theory as proposed by Jaynes \cite{jaynes83}. Whatever
basis we believe, it is plausible that the equilibrium distribution is
the one that maximizes the entropy.

The relativistic equilibrium distribution of a uniform gas
(J{\"u}ttner-Synge distribution) was first obtained by maximizing the
entropy.\cite{juttner11,synge57,tadas09rel-eq} This approach is the
same as the one in this paper, however, energy is the only constraint
in this derivation as usually done in statistical mechanics. Here in
the present paper we generalize it to accommodate the conservation
of angular momentum in addition to energy.

Later, Israel \cite{israel_relativistic_1963} has derived more general
form of equilibrium distribution using kinetic theory. The equilibrium
state is obtained as the distribution with which the collision term
vanishes in the Boltzmann equation. This approach is different from
the one by maximizing entropy; it is pure kinetic and entropy has no
role to play there.  The result includes distributions with
non-uniform local temperature, and it was shown that the gas can be in
equilibrium state when the local inverse temperature four vector
satisfies the Killing equation.

In the present paper we derive the same distribution by maximizing
entropy as done by J{\"u}ttner or Synge \cite{juttner11,synge57}.  The
result includes the non-uniform equilibrium distributions obtained by
Israel\cite{israel_relativistic_1963}. 

\bigskip

Temperature is derived from the Lagrange's coefficient to impose the
constraint of energy conservation in maximizing the entropy.  The same
thing can be done with other kinds of conserved quantities; three
coefficients are derived from three components of momentum, and these
three form a four vector together with the coefficient of energy. This
is called inverse temperature four vector mentioned above. When we
apply the same procedure to angular momentum, the inverse temperature
bi-vector is derived.  We can also obtain the equilibrium velocity
distribution, which is local J{\"u}ttner-Synge distribution moving
along the Killing flow, in agreement with
Israel\cite{israel_relativistic_1963}.

The advantage of this approach is that the role of angular momentum
conservation is clarified as a constraint on the entropy maximization.
Also, we can apply the same approach to systems with gravitation,
in other words, systems in curved spacetimes.  What we investigate
here is the equilibrium state in a flat spacetime, however, the same
thing can be done for static curved spacetimes such as the
Schwarzchild spacetime. If there exist conserved quantities in a given
spacetime, we can calculate the equilibrium state as the maximum
entropy state with the constraints of conservation laws. 

\bigskip{}

There is another advantage of our derivation; the effect of spins can
be incorporated. The result in this paper is obtained with a model of
a classical ideal gas, in which each particle is assumed to be a
volume-less point mass. Therefore, the angular momentum belongs to
orbital motion of particles in this model. We can extend this model to
accommodate intrinsic angular momentum in addition to the orbital
angular momentum.

What carries intrinsic angular momentum in this real world is spins
of particles, and we need quantum physics to analyze it realistically.
However, what we need for our purpose is the conservation of angular
momentum only, and details of quantum aspect is not necessary. Therefore,
we just assume the particles carry intrinsic angular momentum and
do not ask its details in our classical model. 

When we include intrinsic angular momentum, the sum of orbital and
intrinsic angular momentum should be conserved in total. We can calculate
the maximum entropy with the constraint of this total angular momentum
conservation, and resulting equilibrium state includes the thermodynamical
effect of intrinsic angular momentum. Recently Muschik and Borzeszkowski
\cite{muschik_entropy_2009} proposed an thermodynamical expression
of entropy with the effect of spins. Their purpose is to derive a
general form of entropy expression from thermodynamical identity,
and specific form of spin effect is not given. Our result in the present
paper is consistent with theirs, and gives its explicit form by
maximizing entropy.

The conventions in the present paper are the following. Roman indices
for coordinates runs from 1 to 3 and Greek indices runs from 0 to 3,
where 0 represents the temporal coordinate. Four vectors and three
vectors are denoted by indices (e.g., $x_{\mu}$) and bold face (e.g.
$\mathbf{x}$) respectively, and $x_{0}$ is sometimes denoted by $t$ in
non-covariant expressions. The metric tensor is
$\eta_{\mu\nu}=\textnormal{diag}\{1,-1,-1,-1\}$ and the unit system is
such that $c\textnormal{ (speed of light)}=k_{B}\textnormal{ (Boltzman
  constant)}=1.$

\bigskip

Suppose an ideal gas is in a isolated rigid container in
equilibrium. Let $f(t,\mathbf{x},\mathbf{p})$ be a single particle
distribution function, i.e., the probability density to find a
particle near the point $(\mathbf{x},\mathbf{p})=(x_{i},p_{i})$ in six
dimensional position-momentum space at time $t$. Here we
assume the particles in the gas are classical point masses without
internal structure. It should be noted $f$ is a frame-dependent
expression because the time $t$ is along the temporal coordinate axis
of a specific frame. We need complicated form to express the particle
distribution in a covariant way, therefore, we perform calculation
assuming one specific reference frame in the the following. Once the
equilibrium distribution obtained, the distribution function has the
same value in other reference frames, although it is not a fully
covariant expression \cite{vankampen69,tadas09rel-eq}.

What we wish to know is the equilibrium distribution that does not
depend on time $t$, thus we simply denote
$f(\mathbf{x},\mathbf{p})=f(t,\mathbf{x},\mathbf{p})$ in the
following. Note that this expression becomes frame dependent
when the distribution is not spatially uniform as we will discuss
later.  The entropy is defined as\begin{equation}
S=-\int_{V}f(\mathbf{x},\mathbf{p})\ln h^{3}f(\mathbf{x},\mathbf{p})\,
d\mathbf{x}d\mathbf{p}\,,\label{eq:entropy}\end{equation} where $h$ is
the constant in unit of angular momentum to make the argument of
logarithmic function dimensionless. The integration is over the three
volume $V$ of the container, which is the cross section of its world
tube sliced by the surface of $t=\textnormal{constant}$.

We try to
find the equilibrium distribution by maximizing the above entropy
under the constraints of conservation laws. It is known that the
Minkowski spacetime has ten conserved quantities: four components of
energy-momentum and six components of four dimensional angular
momentum. Thus the gas has the following constraints of
energy-momentum conservation,

\begin{eqnarray}
N\int_{V}p^{\mu}f\, d\mathbf{x}d\mathbf{p} & = & P^{\mu}\,,\label{eq:momentum}\end{eqnarray}
and four dimensional angular momentum conservation,\begin{equation}
N\int_{V}(x_{\mu}p^{\nu}-x_{\nu}p^{\mu})f\, d\mathbf{x}d\mathbf{p}=M^{\mu\nu}\,,\label{eq:ang-momentum}\end{equation}
where $P^{\mu}$ , $M^{\mu\nu}$ and $N$ are the total momentum,
angular momentum, and the number of particles respectively. To
calculate (\ref{eq:momentum}) the temporal
component of energy-momentum is given by $p^{0}=\sqrt{m+p_{i}p^{i}}$
with $m$ being the particle rest mass. 

The maximum entropy distribution with constraints (\ref{eq:momentum})
and (\ref{eq:ang-momentum}) can be calculated with the Lagrange's
method as\begin{equation}
f=C\exp(-\beta_{\mu}p^{\mu}-\lambda_{\mu\nu}x^{\mu}p^{\nu})\,,\label{eq:eqdist}\end{equation}
where $\lambda_{\mu\nu}$ is an anti-symmetric tensor (bi-vector)
($\lambda_{\mu\nu}=-\lambda_{\nu\mu}$) and $C$ is a constant normalization
factor to make \begin{equation}
\int_{V}d\mathbf{x}\int_{-\infty}^{\infty}d\mathbf{p}\, f=1\,.\end{equation}
Note that both $\beta_\mu$ and $\lambda_{\mu\nu}$ are the result of
global conservation laws, and thus they do not depend on $x_\mu$. They
can be regarded as thermodynamical parameters similar to the
temperature that characterize the whole system.

The total entropy can be calculated by substituting (\ref{eq:eqdist})
into (\ref{eq:entropy}) as\begin{equation}
S=\beta_{\mu}P^{\mu}+\lambda_{\mu\nu}M^{\mu\nu}-N\ln h^{3}C\label{eq:grbl-ent}\end{equation}
Comparing the above expression with the result of thermodynamics \cite{nakamura_thermodynamics_2010}
we understand that $\beta_{\mu}$ and $\lambda_{\mu\nu}$ are thermodynamical
inverse temperatures.

We can also confirm that the equilibrium distribution of (\ref{eq:eqdist})
agrees with the distribution obtained from the Boltzmann equation
\cite{israel_relativistic_1963} when we introduce the local inverse
temperature four vector $\theta_{\mu}$ as\begin{equation}
\theta_{\mu}(x)=\beta_{\mu}+\lambda_{\mu\nu}x^{\nu}\,.\label{eq:invtemp}\end{equation}
The local velocity $u_{\mu}$ and co-moving temperature (the temperature
measured with a thermometer moving with the local gas velocity) $T$
of the gas at $x$ are then derived as\begin{equation}
T(x)=\frac{1}{\sqrt{\theta_{\mu}\theta^{\mu}}}\,,\:\: u_{\mu}(x)=\frac{\theta_{\mu}(x)}{T(x)}\,.\label{eq:temp-vel}\end{equation}
We can confirm $u_{\mu}$ is actually the local velocity of the gas
by integrating the momentum in the reference frame moving with $u_{\mu}$,
which gives the spatial momentum equal to zero.

\bigskip{}
Now we have the equilibrium distribution (\ref{eq:eqdist}), however,
there was one problem in its derivation. We have started our calculation
by assuming the equilibrium state $f(\mathbf{x},\mathbf{p})$
does not depend on $t=x_{0}$, but $f$ clearly depends on $x_{0}$
when $\lambda_{0i}\ne0$. This is not that $f$ is truly time dependent,
but that $f$ is not spatially uniform and appears to be varying when
measured with spatial coordinates moving relative to the non-uniform
structure. 

The equilibrium state means the state when the system is isolated with
the period long enough. When we say ``time'' it means the timelike
coordinate in a certain coordinate system in relativity. When we
choose an appropriate coordinate system we can obtain a time
independent expression of (\ref{eq:eqdist}).

To obtain time independent distribution, we rotate the three
dimensional coordinates so as to obtain $\lambda_{02}=\lambda_{03}=0$,
which is always possible in general. Then the factor that depends on
time $x^0$ is $\exp[-\lambda_{01}(x^{0}p^{1}-x^{1}p^{0})]$.  When we
introduce pseudo-cylindrical coordinates (Rindler coordinates) for
$x^{0}$ and $x^{1}$ as
\begin{equation} x^{0}=\rho\sinh
a\tau\,,\;\; x^{1}=\rho\cosh a\tau\,.,
\end{equation} 
then we can express the factor that depend on $x^0$ as
\begin{equation}
\exp[-\lambda_{01}(x^{0}p^{1}-x^{1}p^{0})]
= \exp(-\lambda_{\tau\rho} \rho p^\tau)\,,\label{eq:acc-eq}
\end{equation}
which does not depend on our new temporal coordinate $\tau$. 

The Rindler coordinates is known as to represent constant
acceleration; a curve with fixed $\rho$ is the orbit of the constant
acceleration. Therefore, the distribution calculated above is the
equilibrium distribution under the effect of constant acceleration. 
IT should be noted that the above expression does not cover
the entire Minkowski spacetime. It is valid in so called "Rindler
wedges'', in other words, the region where the coordinate $\rho$
becomes timelike; other regions do not have physical interest in
general.

Since the acceleration plays the same role as gravity, the above
distribution has similar structure as the static equilibrium under
constant gravity. The number density becomes smaller towards the
acceleration direction, which is opposite to the equivalent
gravitation. This effect is similar to that of earth's gravity which
makes the atmosphere thinner for higher altitude. When we take
relativistic effect into account, not only the density but also the
local temperature becomes higher towards the direction of
gravity. This is because the energy is equivalent to the mass and
becomes subject to the gravity in relativity.

The same kind of effect takes place in a rotating body with its axis
at rest. It is known in relativistic thermodynamics that a relativistic
wheel has higher local temperature at the outer part; the centrifugal
force plays the role of gravity. This effect can be explained from the
thermodynamical view point, and consistent with our result in
(\ref{eq:eqdist}).

\bigskip

We will examine the effect of intrinsic angular momentum in the following.
We have derived the equilibrium state of an ideal gas based on the
angular momentum conservation above. There the gas
is modeled by classical particles without intrinsic structure. In
contrast, real particles have spins which can contribute to angular
momentum. Quantum theory is required for the realistic description
of spins, however, we can examine qualitative aspects by just assuming
the existence of intrinsic angular momentum without paying attention
to its internal mechanism. For example, one can imagine rotating classical
particles with small enough but finite volumes to mimic three dimensional
intrinsic angular momentum.

When the particles have intrinsic angular momentum, the particle
distribution function $f$ must depend on it as
$f(\mathbf{x},\mathbf{p},m^{\mu\nu})$, where $m^{\mu\nu}$ is the
intrinsic angular momentum of a particle. The constraint of angular momentum
(\ref{eq:ang-momentum}) is then generalized as
\begin{equation}
N\int_{V}(x^{\mu}p^{\nu}-x^{\nu}p^{\mu}+m^{\mu\nu})f\,
d\mathbf{x}d\mathbf{p}=M^{\mu\nu}\,.\label{eq:angular-momentum2}
\end{equation}

Maximizing the entropy (\ref{eq:entropy}) under the above constraint
and (\ref{eq:momentum}) we arrive at\begin{equation}
f=C\exp\left[-\beta_{\mu}p^{\mu}-\lambda_{\mu\nu}\left(x^{\mu}p^{\nu}+\frac{1}{2}m^{\mu\nu}\right)\right]\,.\label{eq:eqdist2}\end{equation}
From the above distribution function we obtain
the total entropy of the gas as\begin{equation}
S=[\beta_{\mu}P^{\mu}+\lambda_{\mu\nu}(\Sigma_O^{\mu\nu}+\Sigma_I^{\mu\nu})-N\ln h^{3}C]\,,\label{eq:gbl-ent2}\end{equation}
where $\Sigma_O^{\mu\nu}$ and $\Sigma_I^{\mu\nu}$ are the 
orbital and intrinsic angular momenta
of the particles, respectively. The spatial density of intrinsic momentum in this
reference frame is obtained by integrating $NV^{-1}m^{\mu\nu}f$ over
$\mathbf{p}$ and $m^{\mu\nu}$. 

The local four velocity of the gas is given by $u_{\mu}$ in (\ref{eq:temp-vel}),
therefore, the flux of intrinsic angular momentum is covariantly expressed
as\begin{equation}
\sigma_{\rho}^{\mu\nu}=\frac{Nu_{\rho}}{V_{0}}\int_{-\infty}^{\infty}m^{\mu\nu}f\, d\mathbf{p}\,dm^{\mu\nu}\,,\end{equation}
where $V_{0}$ is the volume of the container measured in its comoving
frame. The expression for the local entropy flux is then expressed
as \begin{equation}
s_{\rho}=n_{\rho}\ln h^{3}C+\theta_{\mu}T_{\rho}^{\mu}+\frac{1}{2}(\partial_{\nu}\theta_{\mu}-\partial_{\mu}\theta_{\nu})\sigma_{\rho}^{\mu\nu}\label{eq:local-ent2}\end{equation}
where $n_{\rho}$ and $T_{\rho}^{\mu}$ are the particle number flux and
energy-momentum tensor respectively. This expression is confirmed by
integrating it over $V$ with
$\partial_{\nu}\theta_{\mu}-\partial_{\mu}\theta_{\nu}=2\lambda_{\mu\nu}$,
which results in (\ref{eq:gbl-ent2}).

Before closing calculations, we briefly infer the relation of the
above classical toy model to the real quantum spins. In the above
model we treated the intrinsic angular momentum in a classical way,
therefore, it has six continuous components  $m^{\mu\nu}$ in
(\ref{eq:angular-momentum2}).  Consequently, the distribution function
becomes continuous function of $m^{\mu\nu}$ as
$f(\mathbf{x},\mathbf{p},m^{\mu\nu})$.

In contrast, what carries angular momentum in reality is quantum spins
that have discrete values. For example, the kinetic properties of
electrons (spin-$\frac{1}{2}$ particles) are represented by the Wigner
function (quantum distribution function) expressed as
$f_{rs}(\mathbf{x},\mathbf{p})$ with spin indices $r,s=1,2$; the spin
indices represent two spin components in the four dimensional
Minkowski space. Then the local angular momentum of spins
$\sigma_{\rho}^{\mu\nu}$ is given by
\begin{equation}
\sigma_{\rho}^{\mu\nu}=\frac{\hbar}{2}\sum_{s,r}p_{\rho}\xi_{sr}^{\mu\nu}(p)f_{rs}(\mathbf{x},\mathbf{p})\,
d\mathbf{p\,,}\label{eq:spin}
\end{equation}
where 
\begin{equation}
\xi_{s,r}^{\mu\nu}=\frac{1}{4}\bar{u}^{r}(p)\,(\gamma^{\mu\nu}-\gamma^{\nu\mu})u^{r}(p)
\end{equation}
with $u^{r}$(p) and $\gamma^{\mu\nu}$ being normalized spinor
components in momentum space and Dirac matrices, respectively;
$\bar{u}^{r}$ represents Dirac adjoint of
$u^{r}$ (see de Groot et.\ al. \cite{degroot} for detail).

If we can apply our entropy approach in this paper in a
straight-forward manner, we may seek for the equilibrium distribution
by maximizing the entropy $\sum_{r,s}\int f_{rs}\ln
f_{rs}\,d\mbox{\textbf{x}d\textbf{p}}$ under the angular momentum
constraint with spin contribution (\ref{eq:spin}). However, we need
careful examination to validate this procedure. For example, we have
to consider the relation of entropy and the Wigner function $f_{rs}$,
but we do not know how the restriction of energy shell to derive
$f_{rs}$ affects the entropy.  To examine the quantum effects further
is out of the scope of the present paper, and we just point out the
possibility to extend our result to quantum spins here.

\bigskip

In summary, the role of angular momentum in relativistic statistical
mechanics has been investigated in the present paper. The equilibrium
distribution function (\ref{eq:eqdist}) has been derived by maximizing
the entropy under the conservation laws of energy-momentum and angular
momentum.  The distribution obtained is identical to the one obtained
by Israel \cite{israel_relativistic_1963}, however, the stand
points on which equilibrium distributions are derived are quite
different.

Israel \cite{israel_relativistic_1963} obtained the equilibrium state
as the state in which the collision term of the Boltzmann equation
vanishes.  This approach is kinetic and the equilibrium distribution
is obtained as the result of time evolution relaxes to the stationary
state. 

Our approach here, in contrast, is to assume that the equilibrium
state is the one with maximum entropy, in other words, the most
probable state. In this approach, we do not pay attention to the
detailed mechanisms of collisions. What we just assume is the existence
of random energy-momentum/angular momentum exchange process; if it
exist, we can expect the system relaxes to a unique equilibrium state
that does not depend on actual collision mechanisms. 

To examine the angular momentum exchange involving spins, we need to
consider quantum effects for detailed process of collisions. However,
as stated above, we do not need it and our classical toy model here
can provide qualitative insight for the equilibrium with angular
momentum.

\bigskip

The kinetic approach and entropy approach are complimentary and both
have their own advantages. The advantage of the entropy approach is
that the meaning of conservation laws becomes transparent. We have
obtained the equilibrium distribution as the maximum entropy state
under the constraints of energy-momentum and angular momentum
conservation. The effect of angular momentum can be understood in this
process, and we can incorporate intrinsic angular momentum, which is
difficult to deal with the Boltzmann equation.  We have obtained the
entropy expressions with the contribution of intrinsic angular
momentum, (\ref{eq:gbl-ent2}) and (\ref{eq:local-ent2}), consequently.

The derivation of the entropy expressions here based on the assumption
that the gas is in global equilibrium as a whole. However, the differential
expression of (\ref{eq:local-ent2}) may be regarded as the zero-th
order entropy in the dissipative process as usually done in the irreversible
thermodynamics. Actually, it agrees with the one assumed in the past
literature when $\sigma_{\rho}^{\mu\nu}=0$. Further, we can expect
(\ref{eq:local-ent2}) is valid not only for ideal gasses but also
other matters in general as a thermodynamical expression. If intrinsic
angular momentum has the same effect as orbital angular momentum,
the expression for total entropy must be in the form of (\ref{eq:gbl-ent2})
, and then the local entropy (\ref{eq:local-ent2}) can be derived.

Recently a thermodynamical expression of local entropy is derived
including the contribution of intrinsic angular momentum by Muschik
and Borzeszkowski\cite{muschik_entropy_2009}.  Their purpose to obtain
most general entropy expression starting from thermodynamical identity
without any ansatz. Therefore, the result does not give an explicit
form of the entropy contribution.  Our approach here is based on the
ansatz of maximum entropy, and thus able to go one step further to
determine the explicit form.  The contribution of intrinsic angular
momentum is expressed by a linear response matrix in Muschik and
Borzeszkowski\cite{muschik_entropy_2009} and its form agrees with our
expression (\ref{eq:local-ent2}) when the matrix may include
differential operators.

It should be noted, however, there may be some difficulties
when we apply (\ref{eq:local-ent2}) to local equilibrium with
acceleration/gravity. We have obtained time independent expression
(\ref{eq:acc-eq}) by carefully choosing a curved temporal coordinate,
which is possible for a global equilibrium. Such choice of coordinates
may be difficult when each part of gas is in local equilibrium with
different acceleration.

\bigskip

We shall mention one possibility to
experimentally or observationally confirm the results here before
concluding the present paper. We have seen the intrinsic angular
momentum can affect the equilibrium distribution as in
(\ref{eq:eqdist2}). What we have modeled here is an classical ideal
gas with mimicked intrinsic angular momentum, however, this kind of
effect is expected to take place in general. If this effect also
exists in a photon gas, the distribution becomes dependent of the
photon spin when the gas has finite total angular momentum. This means
the black body radiation has dependency on the wave polarization if
the radiation comes from a rotating body. Though we need quantum field
calculation to analyze this effect in detail, a rough estimation can
be made from (\ref{eq:eqdist2}); the effect takes place in the form of
$\exp(-\mbox{\ensuremath{\hbar}}\Omega/T)$ where $\hbar$ is the Planck
constant and $\Omega$ is the angular velocity of rotation.  This may
be an extremely small collection compared to the factor
$\exp(-\mbox{\ensuremath{\hbar}}\omega/T)$ in the Planck distribution
in ordinary environments, however, highly accurate measurement might
be able to detect it.

\end{document}